\documentclass{biophys}
\usepackage{helvet,times}
\usepackage{bm,textcomp}

\jno{kxl014} 
\gridframe{N}
\cropmark{N}

\doi{doi: 10.1529/biophysj.106.090944}

\usepackage[english]{babel}
\usepackage[intlimits]{amsmath}
\usepackage{amssymb}
\usepackage{graphicx}
\usepackage{bm}
\usepackage{amsmath,empheq}
 \usepackage{graphicx}

\usepackage[round,numbers,sort&compress]{natbib}

\global\long\def\koff{k_{\text{off}}}

\global\long\def\kon{k_{\text{on}}}

\begin{document}

\setcounter{page}{1} 

\title{On the Mechanism of Homology Search by RecA Protein Filaments}

\author{M. P. Kochugaeva, A. A. Shvets, A. B. Kolomeisky}

\address{Department of Chemistry, Rice University, Houston, Texas, 77005, USA, 
Center for Theoretical Biological Physics, Rice University, Houston, Texas, 77005, USA}



\begin{abstract}%
{Genetic stability is a key factor in maintaining, survival and reproduction of biological cells. It relies on many processes, but one of the most important is a {\it homologous recombination}, in which the repair of breaks in double-stranded DNA molecules is taking place with a help of several specific proteins. In bacteria this task is accomplished by RecA proteins that are active as nucleoprotein filaments formed on single-stranded segments of DNA. A critical step in the homologous recombination is a search for a corresponding homologous region on DNA, which is called a {\it homology search}. Recent single-molecule experiments clarified some aspects of this process, but its molecular mechanisms remain not well understood. We developed a quantitative theoretical approach to analyze the homology search. It is based on a discrete-state stochastic model that takes into account the most relevant physical-chemical processes in the system. Using a method of first-passage processes, a full dynamic description of the homology search is presented.  It is found that the search dynamics depends on the degree of extension of DNA molecules and on the size of RecA nucleoprotein filaments, in agreement with experimental single-molecule measurements of DNA pairing by RecA proteins. Our theoretical calculations, supported by extensive Monte Carlo computer simulations, provide a molecular description of the mechanisms of the homology search.

}
{Insert Received for publication Date and in final form Date.}
{*Correspondence: tolya@rice.edu.\\
Address reprint requests to Anatoly B. Kolomeisky, Rice University.}
\end{abstract}

\maketitle 

\section*{INTRODUCTION}

A successful functioning of biological cells, which includes accurate DNA replication and transfer of genetic information without errors, strongly depends on stability of its genome \cite{alberts}. However, within a cell DNA molecules constantly experience  attacks by various active chemical molecules, thermal collisions with other molecules and the effect of external radiation \cite{krejci12,renkawitz14, kuzminov01}. This leads to  frequent defects and even breaks in double-stranded DNA chains, which might be lethal for cells.  Fortunately, there is a  natural process known as a {\it homology recombination} that is crucial for repairing DNA double-strand breaks \cite{sagi06,gonda86,renkawitz14,morrical15}. During this process a nucleotide sequence of another identical or similar DNA duplex is used for restoring the original sequence in the broken DNA molecule \cite{chen08,cox07}. Furthermore, the homologous recombination plays a fundamental role in the process of genetic diversity and reassembling of genetic information, which is important for living systems for adaption to changing environments \cite{alberts,gibb12,ragunathan12,sagi06,fulconis06,chen08,renkawitz14}.

In bacteria, a central role in the homologous recombination is played by  RecA proteins \cite{fulconis06,galletto06,heijden08}. They belong to a class of recombination proteins that are  conserved across different organisms, from bacteria to mammals (RadA in archea, and Rad51 and Dmc1 in eukaryotes) \cite{renkawitz14}.  The process of the homologous recombination consists of several stages, and RecA is activated by assembling into a nucleoprotein filament on a single stranded DNA segment that appeared due to a double-strand break in DNA \cite{galletto06,bell12,menetski85,kowalczykowski95,bochkarev97}. One of the  most mysterious  steps in the homology recombination is a process when the RecA nucleoprotein filament is looking  for the corresponding homologous regions on identical undamaged DNA \cite{fu13,pugh88,forget12,morrical15}. This is known as a {\it homology search}. It is still poorly understood, and it remains unknown how RecA nucleoprotein filaments can  find and recognize the homologous sequences among a long DNA chain so quickly and efficiently \cite{renkawitz14,cox07,dorfman04}. By now, it is well proven that the homology search is not an active process since ATP hydrolysis is not required to proceed forward \cite{menetski85,kowalczykowski95}. Therefore, it must be related to protein search for targets on DNA that have been intensively investigated in recent years \cite{hu06,kolomeisky11,kolomeisky12,veksler13,mirny09,koslover11,bauer12,kochugaeva16}. However, a comprehensive description of the mechanisms of the homology search still does not exist even for simplified {\it in vitro} systems. The homology search is much more difficult for understanding for {\it in vivo} conditions  where the identity and functions of various assisting proteins are not fully identified  yet \cite{renkawitz14,cox07}.

Several important experimental advances in studying the homology search have been reported in recent years. Experimental studies that utilized optical trapping and single-molecule fluorescent microscopy were able to visualize how individual RecA nucleoprotein filaments paired broken and undamaged DNA molecules \cite{forget12,gibb12}. It was determined that the homology search strongly depends on the 3D conformational state of the target DNA, i.e.  on the degree of the polymer extension, as well as on the size of the nucleoprotein filament. Based on these observations it was suggested that RecA filament utilizes a so-called {\it intersegmental contact sampling} mechanism, in which multiple contacts with the target DNA are explored \cite{forget12}. As a result, the homologous search is more efficient for longer filaments and for less extended (more coiled) DNA chains. A general question on the role of DNA coiling in the  protein search has been addressed theoretically \cite{hu06,vandenbroek08,lomholt09,leger98}. However, a quantitative model of the intersegment transfer for the homology search has not been developed yet.  

A different idea, which argues that the filament sliding is the source of the high efficiency of the homologous search process, has been also explored recently \cite{ragunathan12}. Single-molecule fluorescent measurements with a high spatiotemporal resolution were employed, which, in contrast to earlier experiments that indicated no sliding \cite{adzuma98}, have observed multiple diffusional events for RecA filaments on DNA. Using computer simulations, it was argued that sliding accelerates the homology search as much as 200 times \cite{ragunathan12}. However, the degree of sliding was not significant: the observed sliding lengths were comparable or less than the size of the RecA filament. In addition, the continuum model utilized in simulations might not be reasonable at these conditions, and more advanced theoretical models show that the sliding actually might not lead to such large accelerations in the search dynamics \cite{veksler13}.

In this paper, we develop a minimalist theoretical model of the homology search that provides a quantitative analysis of the underlying dynamics. It is based on the idea related to the intersegmental contact sampling mechanism, and the model argues that the RecA nucleoprotein filament can scan DNA faster for more coiled DNA conformations. Our analysis extends the earlier developed  discrete-state stochastic approach for protein search for targets on DNA \cite{kolomeisky12,veksler13}, which explicitly takes into consideration major physical and chemical processes in the system. This allows us to obtain a full analytical description for all dynamic properties of the homology search by utilizing a method of first-passage probabilities \cite{veksler13}. Our theoretical calculations, supported by extensive Monte Carlo computer simulations, show that, indeed, extending the DNA chain and/or shortening the RecA filament will slow down the search dynamics. Furthermore, our theoretical model is able to describe quantitatively and explain  the experimental observations, as well as to make testable predictions \cite{forget12}.

\section*{Materials and Methods}

\subsection*{Theoretical Model}

To describe the homology search we introduce a discrete-state stochastic model as presented in Fig. 1. The DNA molecule is viewed as consisting of $L$ binding sites, and one of them (at the site $m$) is the homology segment that the RecA  needs to find and recognize. Thus, the size of each site is equal to the filament's length. The RecA nucleoprotein filament can associate nonspecifically to DNA at any site  with a rate $k_{on}$, and the dissociation rate into the solution is equal to $k_{off}$: see Fig. 1. We divide the bulk volume around DNA into $L$ segments, each of them surrounding the corresponding site on DNA (as shown in Fig. 1). While in the solution, the RecA filament can go from one segment into another segment  with the effective diffusion rate $u$ (Fig. 1). This effectively means that we simplified a three-dimensional motion of the RecA filament into an effective one-dimensional motion. The corresponding diffusion rate depends on the end-to-end extension of the DNA chain: the smaller this distance, the larger is the rate $u$. This assumption explicitly incorporates the idea of the intersegmental contact sampling.  To simplify calculations, a single-molecule view is adopted. We also label all states as $n^{(i)}$ with $i=0$ when the RecA filament is associated to DNA, and $i=1$ when the RecA filament is free in the solution: see Fig. 1.

\begin{figure}
\centering
\includegraphics*[clip,width=0.8\textwidth]{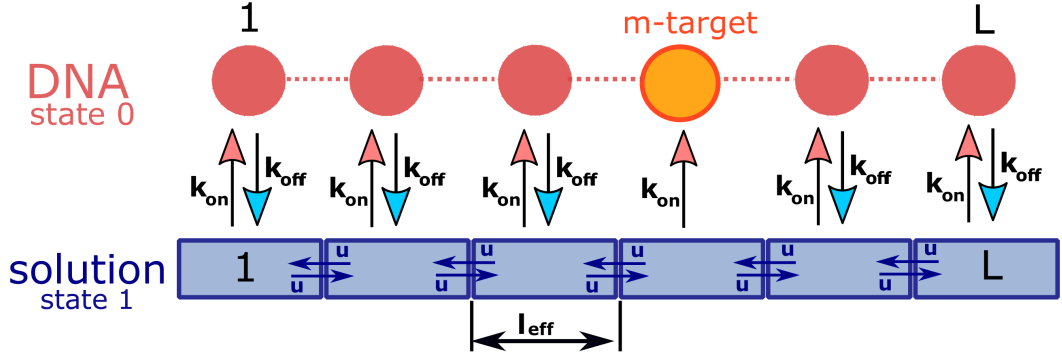}
\caption{A schematic view of the homology search on DNA by RecA filaments. There are $L-1$ nonspecific sites and one specific homology site at the position $m$ on the DNA chain. A filament can diffuse in the solution along the DNA chain with the rate $u$, or might associate to any site on DNA with the rate $k_{on}$. From DNA  the RecA filament can dissociate into the solution with the rate $k_{off}$. The states in the bulk solution are labeled as $1$, and the states on  DNA are labeled as $0$.}
\end{figure}

The RecA filament starts from the solution and the process is completed when it reaches the homology site $m$ for the first time. This suggests to employ a first-passage method of analyzing the dynamic properties, which turned out to be very successful in analyzing various protein search phenomena \cite{kolomeisky12,veksler13,kochugaeva16,esadze14}.  We introduce functions $F_n^{(0)}(t)$ and $F_n^{(1)}(t)$, defined as the probability density functions to reach the target at time $t$ for the first time, if initially at $t=0$ the RecA filament starts at the site $n^{(i)}$ ($n=0, 1, \ldots,L$) on the DNA ($i=0$) or in the solution ($i=1$). The temporal evolution of the first-passage probabilities follow the backward master equations \cite{veksler13}. For $n\ne m$, we have
\begin{equation}\label{me1}
\frac{\partial F_n^{(0)}(t)}{\partial t}=k_{off} F_n^{(1)}(t)-k_{off} F_n^{(0)}(t),
\end{equation}
and 
\begin{equation}\label{me2}
\frac{\partial F_{n}^{(1)}(t)}{\partial t}=u \left[F_{n+1}^{(1)}(t)+F_{n-1}^{(1)}(t)\right]+\kon F_{n}^{(0)}-(2u+\kon)F_{n}^{(1)}(t).
\end{equation}

Because the dynamics is different at the boundaries ($n=1$, $n=m$ or $n=L$), the corresponding  equations are the following:
\begin{empheq}[left=\empheqlbrace]{align}
&\frac{\partial F_{1}^{(1)}(t)}{\partial t}=u F_{2}^{(1)}(t)+\kon F_{1}^{(0)}(t)-(u+\kon)F_{1}^{(1)}(t);\\
&\frac{\partial F_{m}^{(1)}}{\partial t}=u\left[F_{m+1}^{(1)}(t)+F_{m-1}^{(1)}(t)\right]+\kon F_{m}^{(0)}(t)-(2u+\kon)F_{m}^{(1)}(t);\\
&\frac{\partial F_{L}^{(1)}(t)}{\partial t}=u F_{L-1}^{(1)}(t)+\kon F_{L}^{(0)}(t)-(u+\kon)F_{L}^{(1)}(t).
\end{empheq}
In addition, the initial conditions imply that
\begin{equation}\label{initial}
F_{m}^{(0)}(t)=\delta(t).
\end{equation} 
The physical meaning of this expression is the following: if the RecA filaments starts at $t=0$ from the homology sequence site, the search is instantaneously completed.

To determine the first-passage probability functions, we utilize a Laplace transformation, $\widetilde{F}(s)=\intop_{0}^{\infty}e^{-st}F(t)dt$. This allows us to simplify the problem by solving algebraic equations instead of original differential equations. Thus, Eqs. (\ref{me1}) and (\ref{me2}) are modified into
\begin{empheq}[left=\empheqlbrace]{align}
&s\widetilde{F_{n}^{(0)}}(s)=\koff\widetilde{F_{n}^{(1)}}(s)-\koff\widetilde{F_{n}^{(0)}}(s);\label{laplace1}\\
&s\widetilde{F_{n}^{(1)}}(s)=u\left[\widetilde{F_{n+1}^{(1)}}(s)+\widetilde{F_{n-1}^{(1)}}(s)\right]+\kon\widetilde{F_{n}^{(0)}}(s)-(2u+\kon)\widetilde{F_{n}^{(1)}}(s).\label{laplace2}
\end{empheq}
And for the boundaries, taking into account Eq. (\ref{initial}), we have $\widetilde{F_{m}^{(0)}}(s)=1$, the following expressions can be written:
\begin{empheq}[left=\empheqlbrace]{align}
& \left[s+u+\kon-\frac{\kon\koff}{s+\koff}\right]\widetilde{F_{1}^{(1)}}(s)=u\widetilde{F_{2}^{(1)}}(s);\\
& \left[s+u+\kon-\frac{\kon\koff}{s+\koff}\right] \widetilde{F_{L}^{(1)}}(s)=u\widetilde{F_{L-1}^{(1)}}(s);\\
&\left[s+2u+\kon\right]\widetilde{F_{m}^{(1)}}(s)=u\left[\widetilde{F_{m+1}^{(1)}}(s)+\widetilde{F_{m-1}^{(1)}}(s)\right]+\kon.
\end{empheq}
These equations can be solved, producing from Eq. (\ref{laplace1})
\begin{equation}
\widetilde{F_{n}^{(1)}}(s)=\widetilde{F_{n}^{(0)}}(s)\frac{s+\koff}{\koff}.   
\end{equation}
Then from Eq. (\ref{laplace2}), taking into account the boundary conditions, we obtain
\begin{equation}
\widetilde{F_{n}^{(0)}}(s)=A_{1} y^n+A_{2}y^{-n},  
\end{equation}
where
\begin{equation}
y=\frac{s+2u+\kon-\frac{\kon\koff}{s+\koff}-\sqrt{(s+2u+\kon-\frac{\kon\koff}{s+\koff})^2-4u^2}}{2u}; 
\end{equation}
and the coefficients $A_{1}$ and $A_{2}$ have different values depending on the position of the state $n$ with respect to the target. For $n \le m$ we have
\begin{equation}
A_1=\frac{k_{on}(y^m+y^{2L-m+1})}{(\frac{k_{on} k_{off}}{s+k_{off}})(y^m+y^{-m})(y^m+y^{2L-m+1})+u(1-y^2)(1-y^{2L})};
\end{equation}
and
\begin{equation}
A_2=y A_1.
\end{equation}
Similarly, for $n \ge m$ it can be shown that
\begin{equation}
A_1= \frac{k_{on}(y^m+y^{1-m})}{(\frac{k_{on} k_{off}}{s+k_{off}})(y^m+y^{-m})(y^m+y^{2L-m+1})+u(1-y^2)(1-y^{2L})};
\end{equation}
and
\begin{equation}
A_2=y^{2L+1}A_1.
\end{equation}

The knowledge of first-passage distribution functions allows us to obtain a comprehensive description of the dynamics in the system. For example,  we are interested evaluating  the mean search times for the RecA filament to find the homology sequence starting from the solution. It is reasonable to assume that the nucleoprotein filament can start with equal probability at any region of the solution around the DNA chain. Then for the mean search time we derive,
\begin{equation}
T=\frac{1}{L}\sum_{n=1}^{L}\tau_{n}^{(1)}, 
\end{equation}
where 
\begin{equation}
 \tau_{n}^{(1)}=-\frac{\partial \widetilde{F}_{n}^{(1)}}{\partial s}|_{s=0},
\end{equation}
is defined as a mean time to reach the target from the state $n$ in the solution (see Fig. 1). The final expression for the mean search time is given by
\begin{equation}\label{mean-time}
T=\frac{W}{6u \left[\koff/(\kon+\koff)\right]}+\left[\frac{L-1}{\koff}+\frac{L}{\kon}\right],
\end{equation}
 where 
\begin{equation}
W=1+3L+2L^{2}-6m-6Lm+6m^{2}.
\end{equation}
The Eq. (\ref{mean-time}) has a clear physical meaning. The first terms corresponds to the time when the RecA filament diffuses in the solution under condition that it is not bound to the DNA [equal to $\koff/(\kon+\koff)$]. The second term describes the total time for associations and dissociations. So in the limit of very fast diffusion in the solution, $u\rightarrow\infty$, which corresponds to highly coiled DNA configurations, the search time is equal to
\begin{equation}
T=\frac{L-1}{\koff}+\frac{L}{\kon}.
\end{equation}
This expression is easy to understand because on average the filament will make $L-1$ unsuccessful binding attempts to DNA before at the $L$-th step it will find the homology sequence. This also agrees with the analytical result obtained earlier \cite{veksler13}. 

Similar analysis can be done for any dynamic property in the homology search, and we will illustrate this below when the experimental results are analyzed.

\section*{Results and Discussion}

\subsection*{Dynamic Properties of the Homology Search}

The knowledge of first-passage probability functions allows us to fully analyze the dynamics of the  homology search. Several important questions can now be investigated. One of them is the role of the target sequence position along the DNA on the search dynamics. This can be done by varying the target location $m$, evaluating how it affects  the mean search time. The corresponding results are presented in Fig. 2.

One can see that moving the target location along the DNA chain changes the search times for all sets of parameters (see Fig. 2). The fastest dynamics is observed when the homology sequence is in the middle of DNA, and slowest dynamics when the target is at the end. For long DNA chains, $L \gg 1$, it is up to 4 times faster to find the target if it is in the center versus the end of the DNA molecule (Fig. 2).  This can be easily explained by noting that to reach the homology sequence the RecA filament must first to come to the volume segment around the target. Because the filament starts with equal probability anywhere in the volume around the DNA chain, the dynamics is faster if the target is in the middle of the chain. At average, the protein filament must diffuse $\sim L/4$ segments to reach the segment around the target in the middle, while for the end target locations it will move through  $\sim L/2$ volume segments. The time diffuse in the solution scales quadratically with the distance. Then the search slowing down due to moving the target to the end is given by $T/T_{max} \simeq \frac{(L/4)^{2}}{(L/2)^{2}}=1/4$ (see Fig. 2). 

It is interesting to note that this behavior differs for other protein search systems where three-dimensional diffusion is usually very fast \cite{veksler13,kochugaeva16}.  For these systems, the relative  contribution of the sliding along DNA in comparison with the bulk diffusion determines the importance of the varying the target position. If the search is dominated by the motion via the bulk solution, the variation of the target position is less relevant for the protein search dynamics.

\begin{figure}
\centering
\includegraphics*[clip,width=0.6\textwidth]{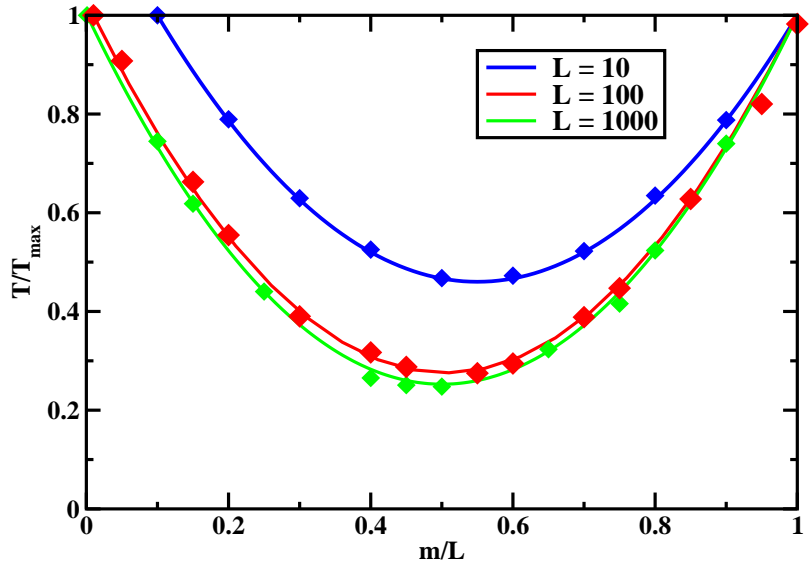}
\caption{Normalized mean search time to find the homology sequence as a function of the relative target position for different DNA lengths. Solid lines represent theoretical predictions, and symbols correspond to Monte-Carlo simulations. Parameters used for calculations are: $\kon=u=10^5$ s$^{-1}$ and $\koff=10^3$ s$^{-1}$.}
\end{figure}

The next question is to understand the effect of the length of DNA in the homology search. The dependence of the search time on $L$ is presented in Fig. 3. Two different behaviors are observed. When the protein filament spends most of the time in the solution ($u \ll \kon$), the mean search time has a quadratic scaling, $T \sim L^{2}$. In the solution, the RecA protein filament performs a simple unbiased diffusion. At the same time, when the rate limiting step is associated with binding/unbinding events to/from DNA ($u \gg \kon, \koff$), the scaling is linear, $T \sim L$ because the filament visits each site on DNA independently of each other. The same  conclusions can be obtained from Eq. (\ref{mean-time}). Thus, for RecA to conduct fast and efficient search it should not stay very long in the solution if DNA chains are quite long. This also means that the search is generally faster for longer RecA filaments because a smaller number of associations to DNA is needed to find the homology sequence.

\begin{figure}
\centering
\includegraphics*[clip,width=0.6\textwidth]{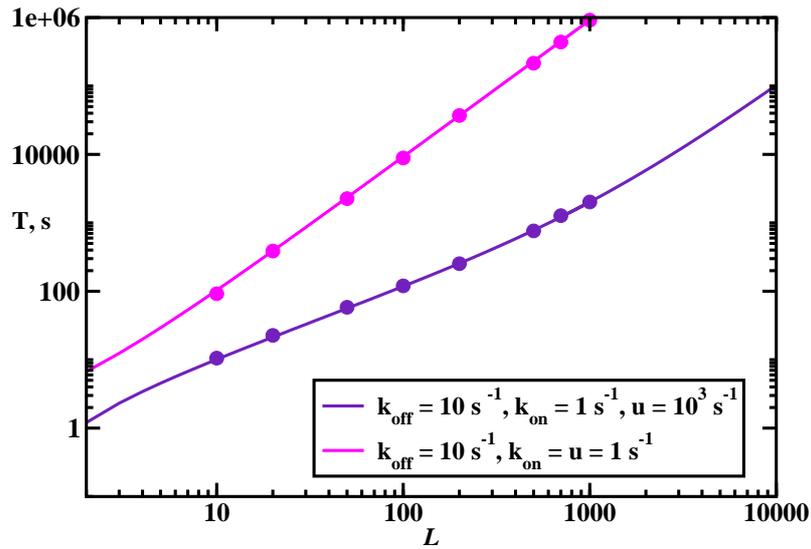}
\caption{Mean search times to find the homology sequence as a function of the DNA length. Solid lines represent theoretical predictions, and symbols correspond to Monte-Carlo simulations. The homology sequence is in the middle of the DNA chain ($m=L/2$).}
\end{figure}

Another important factor in the homology search is the strength of protein-DNA interactions. To quantify the affinity between protein and DNA molecules, we consider an equilibrium binding constant $K=\kon/\koff$ that specifies the tendency of the protein to associate non-specifically to DNA. The mean search times as a function of the binding affinity are illustrated in Fig. 4. A non-monotonic behavior is observed, which can be explained using the following arguments. For weak affinities, $K \ll 1$ the protein filament does not like to associate to DNA, and this prevents it from quickly finding the homology sequence. For strong affinities, $K \gg 1$, the effect is opposite. The non-specific interaction between the RecA filament and DNA is so strong that the protein is frequently trapped at different locations on DNA. This again slows down the search dynamics. Only for intermediate affinities, $K \sim 1-10$ the homology search  is fast because the protein can efficiently scan the DNA without being trapped. Experiments suggest that RecA operate in this regime of affinities \cite{forget12}.

\begin{figure}
\centering
\includegraphics*[clip,width=0.6\textwidth]{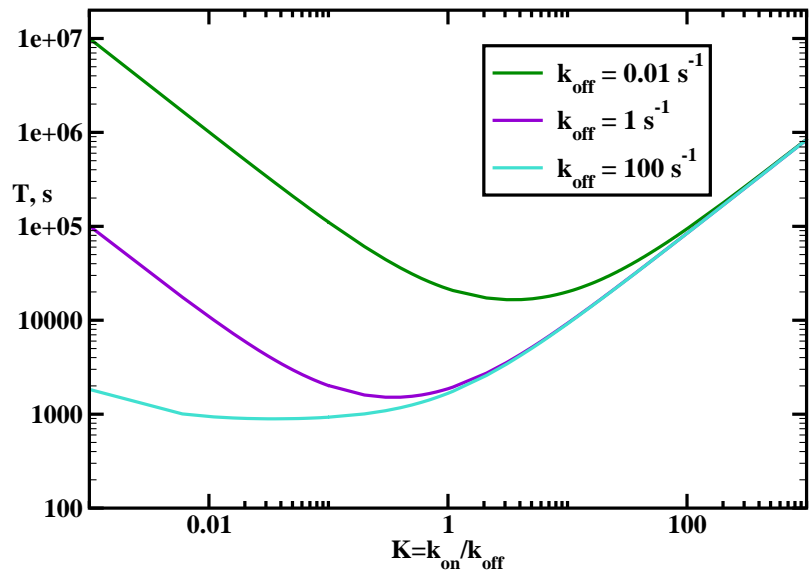}
\caption{Mean search times to find the homology sequence as a function of the protein-DNA affinity, $K=\kon/\koff$. Different curves correspond to different but fixed dissociation transition rates $\koff$. Parameters used for calculations are: $L=100 l$, $m=50 l$ and $u=1$ s$^{-1}$.}
\end{figure}

Our theoretical approach allows us to fully describe all possible search behaviors in the system. To show this we build a dynamic phase diagram for the homology search, as shown in Fig. 5. To understand the dynamics we note that there are two relevant length scales in the system. The first one is the length of the DNA chain, $L$. The second one is a length $d=\sqrt{\frac{u}{\kon}}$, which has a physical meaning of the average distance that the RecA filament diffuses in the solution before binding to DNA. These two length scales lead to two different dynamic search regimes: see Fig. 5. When $d \ll L$, the nucleoprotein filament has a strong tendency to non-specifically bind DNA at any position, and this effectively traps the protein on DNA. The search time decreases with increasing the length $d$ because it corresponds to relaxing the trapping on DNA, leading to faster search for the homology sequence. The situation is different for $d \gg L$, when the protein filament prefers to be found in the solution. But this does not help with the search. Increasing the length $d$  also increases the mean search time, and this makes the search is even less efficient. One can clearly see that the most optimal search is observed when the protein diffusion in the solution is balanced by frequent associations and dissociation from DNA.

It is interesting to note that this dynamic phase diagram is also different from other protein search systems, where usually three dynamic regimes are observed \cite{veksler13,kochugaeva16}. The main reason for this is the absence of the sliding along the DNA, which is associated with another length scale and a different dynamic regime.

\begin{figure}
\centering
\includegraphics*[clip,width=0.6\textwidth]{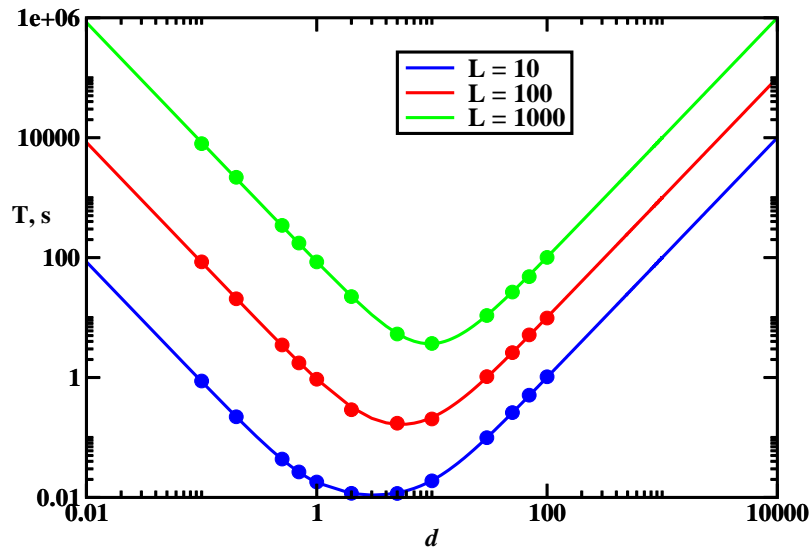}
\caption{Mean search times to find the homology sequence as a function of the characteristic length $d$ for different DNA lengths. Solid curves correspond to theoretical predictions, while symbols are from Monte Carlo computer simulations. Parameters used for calculations are: $m=L/2$, $u=10^5$ s$^{-1}$ and  $\koff=10^3$ s$^{-1}$.}
\end{figure}

\section*{Comparison with Experiments}

One of the advantages of our theoretical approach is the fact that it provides a fully analytical description of the homology search dynamics. It can be tested by applying it for a quantitative description of the experiments on RecA. Here we present the analysis of the single-molecule observations of {\it in vitro} RecA homology search \cite{forget12}. In these experiments, the degree of DNA pairing by RecA filaments has been measured for different end-to-end distances in DNA and for different filaments lengths \cite{forget12}.

To assist our analysis, we first have to evaluate the effective diffusion rate $u$. Because the RecA nucleoprotein filaments utilized in these experiments usually are not very long (several hundreds of nucleotides in length), they can be viewed  as rigid cylindrical tubes. Then the translational diffusion constant for the cylindrical object of length $l$ and diameter $a$ in the solution of viscosity $\eta$  can be written as \cite{ortega03}  
\begin{equation}
 D=\frac{k_{B}T\left[\ln(l/a)+c \right]}{6 \pi \eta l},  
\end{equation}
where a numerical parameter $c$ is a finite-size correction. For long cylinders, $l \gg a$, it was numerically evaluated that $c \simeq 0.312$ \cite{ortega03}. For RecA protein the diameter can be estimated as $a \simeq 2$ nm \cite{xing04}, and the filament with 500 nucleotides has a length $l \simeq 150$ nm, so that $l$ is always much larger than $a$. Assuming that the viscosity of the solution can be reasonably approximated as the viscosity of water, $\eta(water)=8.9\cdot10^{-4}$ kg$\cdot$ m/s, and considering the RecA nucleoprotein filaments of the length of ~500 nucleotides, we estimate the translational diffusion coefficient as  $D \simeq 7 \times 10^{-12}$ m$^{2}$/s. 

To obtain the estimate for the effective diffusion rate $u$, we note that the volume around the DNA molecule is divided into $L/l$ segments, and the effective one-dimensional size of each segment is given by $l_{eff}=R/(L/l)$, where $R$ is the end-to-end DNA distance (see also Fig. 1). It takes the average time $1/u$ to move the distance $l_{eff}$ in our model.  Simultaneously, the time for the RecA filament to diffuse the same distance in the solution is $l_{eff}^{2}/2D$. Then the condition,
\begin{equation}
\frac{1}{u} \simeq \frac{l_{eff}^{2}}{2D}=\frac{2 D^{2} L^{2}}{l^{2} R^{2}},
\end{equation}   
leads to the explicit estimate for the rate $u$. For experimental parameters that describe the DNA pairing by RecA filaments ($L=48502$ bps) \cite{forget12}, it can be shown that $u \simeq 10^{3}$ s$^{-1}$ for the fully extended DNA chains ($R = 16$ $\mu$m), while for the coiled DNA chains ($R = 2$ $\mu$m) we obtain $u \simeq 10^{5}$ s$^{-1}$. Experiments also suggest that for the filament concentration of 100 pM, the equilibrium binding constant is $K >10$, giving the connection between the transition rates $\kon$ and $\koff$ \cite{forget12}.

The experimentally observed fraction $f(t)$ of RecA molecules that find the homology sequence at the time $t$  can be evaluated from the first passage probability functions, averaged over the initial starting conditions,
\begin{equation}
f(t)=\frac{1}{L} \sum_{n=1}^{L} \int_{0}^{t} F_{n}^{(1)}(t) dt.
\end{equation}
These quantities are explicitly calculated by numerically inverting the Laplace transforms $\widetilde{F_{n}^{(1)}(s)}$ using the procedure described in Ref. \cite{valko04}. Experimental measurements on the degree of DNA pairing by RecA filaments are described then with only one fitting parameter, and the results are presented in Figs. 6 and 7. 

As one can see from Fig. 6, extending the distance between DNA ends makes the homology search less efficient. When  DNA is coiled (2 $\mu$m end-to-end distance) more than 85$\%$ of the targets are found in less than 2 minutes. For the extended DNA chain (8 $\mu$m end-to-end distance) less than 10$\%$ of the homology sequences are located during the same time period.  Similar behavior is presented in Fig. 7 where the temporal evolution of the fraction of the successful homology search events is presented. Our theoretical picture, which argues for longer search times for more extended DNA configurations, is able to capture these experimental observations reasonably well.

\begin{figure}
\centering
\includegraphics*[clip,width=0.6\textwidth]{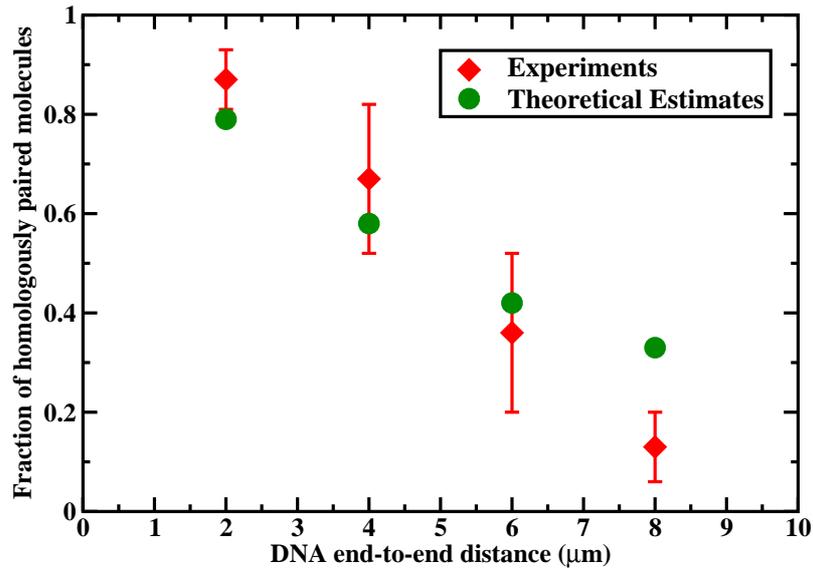}
\caption{Fraction of the homologously paired RecA molecules as a function of the DNA end-to-end distance for $l=430$ nucleotides filaments  2 minutes after the beginning of the homology search. Red symbols are experimental data from Ref. \cite{forget12}, and green circles are theoretical estimates. The following parameters were utilized for calculations: $\kon=1000$ s$^{-1}$, $\koff=2$ s$^{-1}$,  $m= 55l$, $L = 113  l$; the effective diffusion rates $u=24536$ s$^{-1}$, 6134 s$^{-1}$, 2726 s$^{-1}$ and 1533 s$^{-1}$ for $R=2$$\mu$m, 4$\mu$m, 6$\mu$m  and 8$\mu$m DNA end-to end distances, respectively.} 
\end{figure}

\begin{figure}
\centering
\includegraphics*[clip,width=0.6\textwidth]{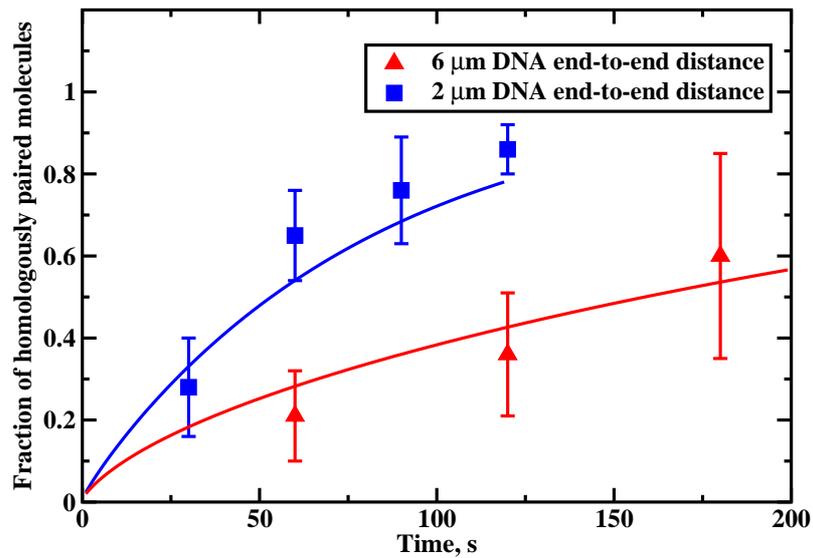}
\caption{Fraction of the homologously paired RecA molecules as a function of the time for 430 nucleotides filaments. Symbols are experimental data from Ref. \cite{forget12}, and solid curves are theoretical predictions. The following parameters were utilized for calculations: $\kon=1000$ s$^{-1}$, $\koff=2$ s$^{-1}$,  $m= 55$, $L = 113$; the effective diffusion rates $u=24536$ and 2726 s$^{-1}$ for 2 and 6 $\mu$m DNA end-to end distances, respectively.}
\end{figure}

Because of the fully quantitative nature of our theoretical model, not only existing experimental observations can be described. We can also make specific predictions that can tested in lab. More specifically, for the experimental conditions described in Ref. \cite{forget12}, the dependence of the search time on the degree of coiling of DNA molecules is presented in Fig. 8. We predict that the homology search will be accomplished in less than 2 minutes for coiled DNA chains, while for the extended conformations it might take up to 8-10 minutes.

\begin{figure}
\centering
\includegraphics*[clip,width=0.6\textwidth]{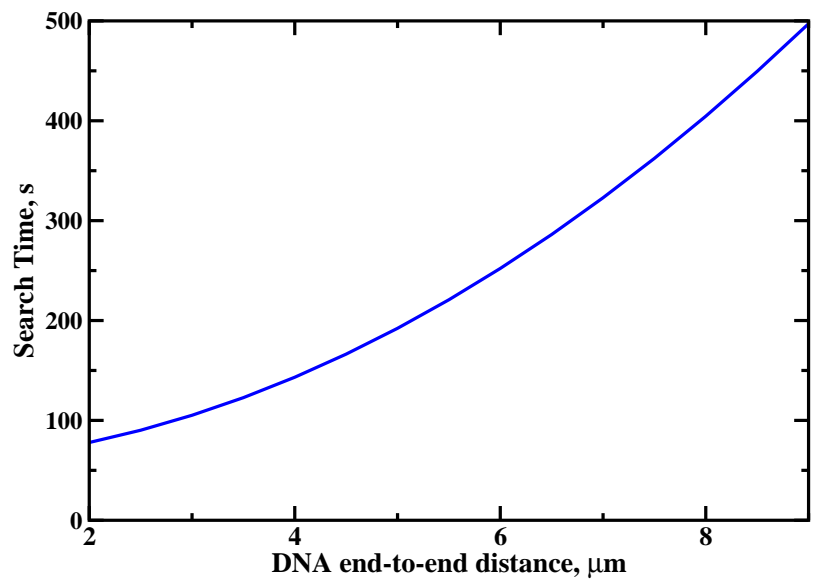}
\caption{Theoretical predictions for homology search times as a function of the DNA end-to-end distance for experimental conditions in Ref. \cite{forget12}.}
\end{figure}

\section*{Conclusion}

We developed a fully quantitative theoretical approach to describe the homology search process by RecA nucleoprotein filaments. It is based on the idea of intersegment sampling, which suggests that it takes longer to probe more extended DNA conformations and for shorter filaments. Our discrete-state stochastic model of the homology search, which takes into account the most relevant physical-chemical processes in the system, is solved analytically using the method of first-passage processes. 

The presented theoretical analysis shows that the location of the homology sequence influence the search dynamics: the search is faster up to 4 times if the target is in the middle of the DNA chain in comparison with the location at DNA ends. We also found that the dependence of the search time on the DNA length is determined by dominating process in the system. If the RecA filament spends most of the time in the solution around the DNA molecule, then the search time has a quadratic dependence on the DNA length. However, if association to DNA and dissociation from DNA are rate-limiting steps, then the linear scaling is observed. In addition, the non-specific protein-DNA interactions have been identified as another factor that affects the homology search. Our theoretical calculations suggest that there is the optimal protein-DNA affinity that speeds up the search dynamics.  It was argued that this effect is the result of the compromise between being trapped on DNA for strong attractive interactions and not coming to DNA at all for strong repulsions. At the end, we constructed a dynamic phase diagram for the homology search, where two different regimes were found. When the distance traveled by the filament in the solution is much smaller than the DNA length, the RecA molecule has a strong tendency to be trapped on DNA. Increasing this length accelerates the search. However, when this length becomes larger than the DNA length, the trend is reversed because the filament is mostly in the solution, and this increases their search times for the homology sequence. Furthermore, theoretical description of the homology search indicates that it differs in many aspects from other processes of proteins searching for targets on DNA. Finally, our theoretical picture is successfully applied for analyzing single-molecule experiments on DNA pairing by RecA filaments. We are able to quantitatively account for the effect of DNA coiling and size of RecA filaments during the homology search. Experimentally testable predictions on the homology search times are also presented.

Although our theoretical approach provides a simple quantitative description of the homology search, which also agrees reasonably well with single-molecule experimental measurements, there are several issues that should be discussed. It has been argued that the intersegmental contact sampling is the leading mechanisms in the homology search \cite{forget12}. However, our theoretical method only partially takes this mechanism into account. We incorporate the idea that it is longer to search more extended DNA chains, but we implicitly assume that the RecA protein binds and dissociates as a whole filament during the search process. In reality,  some parts of the filament can dissociate from DNA, simultaneously binding to other segments on DNA. This possibility of intersegment transfer is not accounted for in our method. These arguments suggests that our model should work better for more extended DNA conformations, and its accuracy for more coiled DNA configurations is less reliable. At the same time, one should notice that the intersegment transfer has been successfully accounted in other protein search systems \cite{esadze14}, so our theoretical method can be extended in this direction. Another important process that was not taken into account in our theoretical model is the sliding of RecA filaments along the DNA chains \cite{ragunathan12}.  It is still unclear if this effect is relevant for the homology search, but the proposed discrete-state stochastic framework can be extended to take care of this question. Despite these issues, we believe that our method captures some important physical-chemical aspects of the mechanisms of the homology search. It will be important to test the theoretical predictions in experimental studies, as well as in more advanced theoretical treatments.

\section*{Acknowledgments}

The work was supported by the Welch Foundation (Grant C-1559), by the NSF (Grant CHE-1360979), and by the Center for Theoretical Biological Physics sponsored by the NSF (Grant PHY-1427654). We also would like to acknowledge the help of Dr. Peter Valko for clarifying some technical issues with numerical calculations.










\end{document}